\documentclass{mem}
\usepackage{natbib}\usepackage{txfonts}\usepackage{balance}
\usepackage{graphicx}
\usepackage[a4paper,breaklinks,dvipdfm]{hyperref}
\idline{75}{282}
\begin{document}

\title{Terzan 5: a Fossil Remnant of the Galactic Bulge}

\subtitle{}

\author{
D. \,Massari\inst{1} 
\and F.R. \, Ferraro\inst{1}
\and E. \, Dalessandro\inst{1}
\and B. \, Lanzoni\inst{1}
\and A. \, Mucciarelli\inst{1}
\and L. \, Origlia\inst{2}
          }

  \offprints{D. Massari}

\institute{
Universit\`{a} di Bologna, Dipartimento di Fisica e Astronomia,
Viale Berti-Pichat 6/2, 40127, Bologna, Italy
\and
INAF-Osservatorio Astronomico di Bologna,
Via Ranzani 1, 40127, Bologna, Italy\\
\email{davide.massari@unibo.it}
}

\authorrunning{Massari }

\titlerunning{Terzan 5, a Fossil Remnant of the Galactic Bulge}

\abstract{
Terzan 5 is a stellar system located in the Galactic Bulge, at a distance of 5.9 kpc. 
Recent discoveries show that it hosts two stellar populations with different iron abundance ($\Delta$[Fe/H]=0.5). 
Such a large difference has been measured only in $\omega$ Centauri in the Galactic halo. 
Moreover no anticorrelation is observed in Terzan 5, hence it is not a genuine globular cluster.
The observed chemical patterns are strikingly similar to those observed in the Bulge stars. This suggests that
Terzan 5 is a remnant fragment of the Galactic bulge.
}
\maketitle{}

\section{General framework} 

\begin{figure}[!htp]
\resizebox{\hsize}{!}{\includegraphics[clip=true]{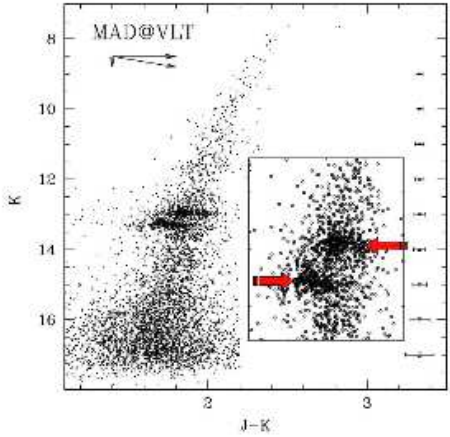}}
\caption{
\footnotesize
The two RCs of Terzan 5. In the main panel,
the (K, J-K) CMD of the central region of Terzan 5. In the inset,
a magnified view of the RC region, with the two RCs marked
with (red in the online version) arrows. Error bars are also plotted at different magnitude
levels. For details see \cite{f09}.
}
\label{hbs}
\end{figure}

\begin{figure}[!htp]
\resizebox{\hsize}{!}{\includegraphics[clip=true]{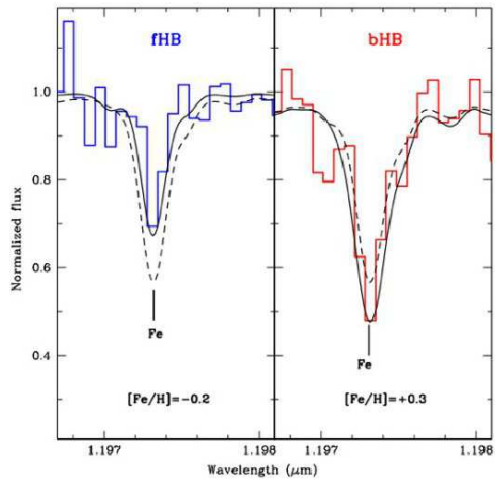}}
\caption{
\footnotesize
J-band spectra near the 1.1973 $\mu$m iron line for faint RC (left)
and bright RC (right) stars. The black solid lines correspond to the
best fit synthetic spectra obtained for temperatures and gravities
derived from evolutionary models reproducing the observed colours of 
the RCs stars (for details see \citealt{f09}).
}
\label{irons}
\end{figure}

Terzan 5 is a stellar system commonly catalogued as a globular cluster
(GC), located in the inner bulge of our Galaxy, at a distance of 5.9 Kpc \citep{valenti}. 
It also hosts an exceptionally large population of millisecond Pulsars (MSPs).
Indeed the 34 MSPs detected so far in Terzan 5 amount to about the 25\%
of the entire sample of known MSPs in Galactic Globular Clasters (GCs, see \citealt{ransom}).
By using a set of high-resolution images in K and J bands taken with the Multi-conjugate
Adaptive Optics Demonstrator (MAD),
\cite{f09} discovered the presence of two distinct sub-populations,
which define two red clumps (RCs) clearly separated in luminosity in the
$(K,J-K)$ color-magnitude diagram (CMD, see Fig. \ref{hbs}).
A prompt spectroscopic follow up performed with the near-IR spectrograph (NIRSPEC) 
mounted at the Keck II Telescope demonstrated that the two populations
have the same radial velocities (hence they belong to the same stellar system)
and they show significantly
different iron content: the bright RC at $K = 12.85$ is
populated by a quite metal rich (MR) component ([Fe/H]$\simeq +0.3$),
while the faint clump at $K = 13.15$ corresponds to a relatively metal
poor (MP) population at [Fe/H]$\simeq -0.2$ (Fig. \ref{irons}). 
These findings 
confirm the existance of two distinct stellar populations in Terzan 5
and the comparison with theoretical stellar isochrones 
suggests that they possibly have been generated by two bursts of
star formation separated by a few ($\sim 6$) Gyr.
While the age gap can be reduced by invoking a difference in the helium content
of the two populations (\citealt{dantona}), the iron enrichment and the spatial
segregation of the brightest clump, together with the extraordinary amount of MSPs
found in Terzan 5, indicate that this system probably experienced a particularly troubled
formation and evolution.
\textit{Terzan 5 is the first GC-like system in the Galactic bulge found to have a spread
in the iron content}.
Before this discovery,
such a large difference in the iron content ($\Delta$ [Fe/H]$>0.5$
dex) was found only in $\omega$ Centauri, a GC-like system in
the Galactic halo, which is believed
to be the remnant of a dwarf galaxy accreted by the Milky Way.

\begin{figure}[!htp]
\resizebox{\hsize}{!}{\includegraphics[clip=true]{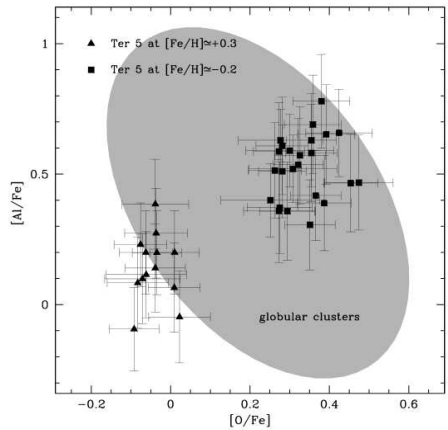}}
\caption{
\footnotesize
[Al/Fe] vs. [O/Fe] abundance ratios of the Terzan 5 giants observed by \cite{origlia}.
The grey ellipse indicates the range of values measured in Galactic GCs.
}
\label{antico}
\end{figure}

\begin{figure}[!htp]
\resizebox{\hsize}{!}{\includegraphics[clip=true]{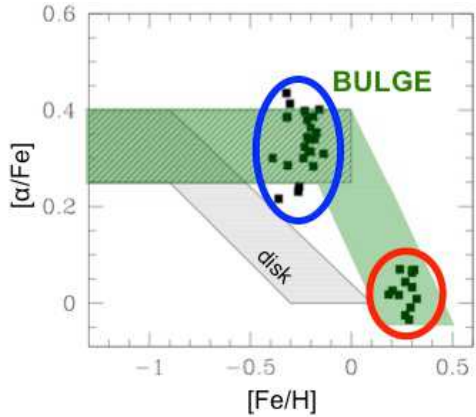}}
\caption{
\footnotesize
[$\alpha$/Fe] vs. [Fe/H] abundance ratios for Terzan 5 giants. The behavior
of the two populations follows that of the bulge stars (green region),
suggesting a strong evolutionay link between Terzan 5 and the bulge itself.
}
\label{alfafe}
\end{figure}

\cite{origlia} presented a detailed study of the abundance patterns of
Terzan 5. First of all, their study demonstrated that the abundances of light elements
(like O, Mg, and Al) measured in both the sub-populations do not
follow the typical anti-correlations observed in genuine GCs (see \citealt{carretta}),
neither in the population as a whole, nor in the single ones (Fig. \ref{antico}).
Secondly the
overall iron abundance and the $\alpha-$enhancement of the MP
component demonstrate that it formed from a gas mainly enriched by
Type II supernovae (SNII) on a short timescale, while the progenitor
gas of the MR component was further polluted by SNIa on longer
timescales. 
These chemical patterns are strikingly similar to
those measured in the bulge field stars as shown in Fig. \ref{alfafe}.

These observational results demonstrate that Terzan 5 is not a genuine
GC, but a stellar system that has experienced complex star formation and
chemical enrichment histories. Indeed it is likely to have been much more
massive in the past than today (with a mass of at least a few
$10^7-10^8 M_\odot$, while its current value is $\sim 10^6
M_\odot$; \citealt{l10}), thus to retain the high-velocity gas ejected
by violent SN explosions. Moreover the collected evidence indicates
that it formed and evolved in strict connection with its present-day
environment (the bulge)\footnote{The probability that Terzan 5 was
  accreted from outside the Milky Way (as supposed for $\omega$
  Centauri) is therefore quite low.}, thus suggesting the possibility
that it is the relic of one of the pristine fragments that contributed
to form the Galactic bulge itself.  In this context, also the
extraordinary population of millisecond pulsars (MSPs) observed in
Terzan 5 can find a
natural explanation. In fact, the large number of SNII required to
account for the observed abundance patterns would be expected to have produced
a large population of neutron stars, mostly retained by the deep
potential well of the massive {\it proto}-Terzan 5.  
In addition, the large
collisional rate of this system (\citealt{vhut}, \citealt{l10}) may also have favored
the formation of binary systems containing neutron stars and promoted
the re-cycling process responsible for the production of the large MSP
population now observed in Terzan 5. 

\section{The project}

\begin{figure}[!htp]
\includegraphics[height=6cm,width=6.5cm]{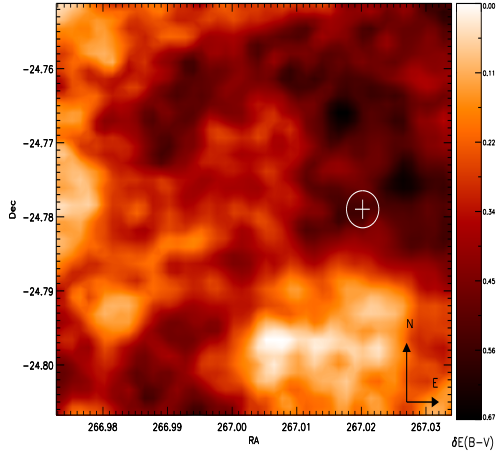}
\caption{
\footnotesize
 Reddening map
   ($200\arcsec\times200\arcsec$) in the direction of Terzan 5. The
   differential color excess $\delta E(B-V)$ ranges between zero (lightest)
   and 0.67 (darkest).  The gravity center and core
   radius of Terzan 5 (\citealt{l10}) are marked for reference as
   white cross and circle, respectively. See \cite{massari} for details.
}
\label{map}
\end{figure}

\begin{figure}[!htp]
\resizebox{\hsize}{!}{\includegraphics[clip=true]{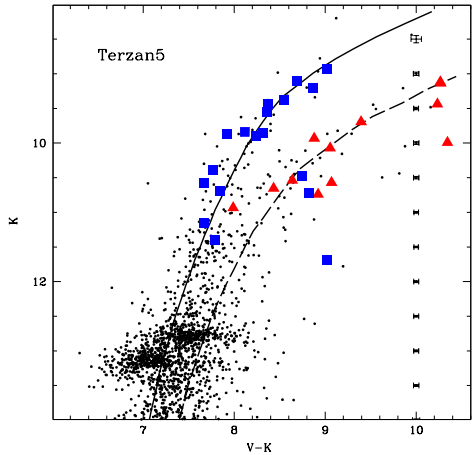}}
\caption{
\footnotesize
Brightest portion of the differential reddening corrected
$(K,V-K)$ CMD of Terzan 5, with error bars also reported.  
Beside the two RCs, also two well separated RGBs are clearly
distinguishable. The solid and dashed lines correspond to the mean
ridge lines of the MP and the MR sub-populations, respectively.
Stars with $[Fe/H]<0$ have been overplotted as squares (blue in the
online version), while those with $[Fe/H]>0$ as (red) triangles 
(abundances are taken from \citealt{origlia}).
}
\label{rgbs}
\end{figure}

Within this exciting scenario, we are now conducting a project aimed
at reconstructing the origin and the evolutionary history of Terzan 5
and on a larger scale of the Galactic bulge, by looking for other systems
similar to Terzan 5.
Since the locations of the two RCs can be due to a proper combination of 
different ages and He content (\citealt{f09}, \citealt{dantona}), an accurate estimate of the 
absolute ages is urged via the measure of the Turn Off luminosity.
However, severe limitations to the detailed analysis of the
evolutionary sequences are introduced by the
presence of large differential reddening. To face this problem \cite{massari}
built the highest-resolution extinction map ever constructed in the
direction of Terzan 5. The differential extinction,
measured with a spatial resolution of $8\arcsec \times
8\arcsec$, turned out to vary from $E(B-V)=2.15$ mag, up to 2.82 mag
(see Fig. \ref{map}).
A free tool providing the color excess values at any coordinate within the map
can be found at the Web site {\tt http://www.cosmic-lab.eu/Cosmic-Lab/}.
After the correction
for differential reddening, two distinct red giant branches become
clearly visible in the color magnitude diagram of Terzan 5 and they
well correspond to the two sub-populations with different iron
abundances recently discovered in this system (Fig. \ref{rgbs}).

10 Hubble Space Telescope (HST) orbits have been allocated for observations with 
the Wide Field Camera 3 (WFC3) in the F110W ($\sim$ J) and F160W ($\sim$ H) filters
in the current HST Cycle 20.
In addition, we obtain second-epoch optical observations with the Advanced Camera for
Surveys (ACS) aboard the HST that will be combined with similar archival images
to perform a detailed relative proper motion analysis to clean
the CMDs from non-member components and give a final answer to the 
star fomation history of this system. 
Also, we are performing a detailed screening of the abundances of Terzan 5 through
the analysis of hundreds stars observed with the spectrograph XSHOOTER, mounted 
at the European Southern Observatory Very Large Telescope.

Finally a total of 4 nights have been assigned to our group for a spectroscopical
study with XSHOOTER with the aim of looking for other bulge GC-like system
that experienced a complex star formation history similar to that of Terzan 5.

\begin{acknowledgements}
This research is part of the project COSMIC-LAB
funded by the European Research Council (under contract
ERC-2010-AdG-267675).
\end{acknowledgements}

\bibliographystyle{aa}

\end{document}